# Experimental Demonstrations of Native Implementation of Boolean Logic Hamiltonian in a Superconducting Quantum Annealer

Daisuke Saida[1], Yuki Yamanashi[2], Mutsuo Hidaka[1], Fuminori Hirayama[1], Kentaro Imafuku[1], Shuichi Nagasawa[1] and Shiro Kawabata[1]

[1]National Institute of Advanced Industrial Science and Technology, 1-1-1 Umezono, Tsukuba, Ibaraki, Japan

[2]Yokohama National University, 79-5 Tokiwadai, Hodogaya-ku, Yokohama, Kanagawa, Japan

Corresponding author: Daisuke Saida (email: saida.daisuke@aist.go.jp).

**ABSTRACT** Experimental demonstrations of quantum annealing with "native" implementation of Boolean logic Hamiltonians are reported. As a superconducting integrated circuit, a problem Hamiltonian whose set of ground states is consistent with a given truth table is implemented for quantum annealing with no redundant qubits. As examples of the truth table, NAND and NOR are successfully fabricated as an identical circuit. Similarly, a native implementation of a multiplier comprising six superconducting flux qubits is also demonstrated. These native implementations of Hamiltonians consistent with Boolean logic provide an efficient and scalable way of applying annealing computation to so-called circuit satisfiability problems that aim to find a set of inputs consistent with a given output over any Boolean logic functions, especially those like factorization through a multiplier Hamiltonian. A proof-of-concept demonstration of a hybrid computing architecture for domain-specific quantum computing is described.

## I. INTRODUCTION

Quantum annealing (QA) is a mechanism for solving combinatorial optimization problems [1-4]. Physical implementations of mechanisms utilizing inductively coupled superconducting qubit systems are being actively investigated [5-14]. As a prime example, commercial quantum annealers from D-Wave Systems provide opportunities to map optimization problems into a 5000-qubit system [15]. Their systems utilize superconducting flux qubits [5,6,10]. To provide general versatility, they employ a chimera graph architecture composed of unit cells with eight qubits and tunable couplers [8].

As a general idea of QA, quantum dynamics governed by the following Hamiltonian is considered:

$$H = -\frac{A(t)}{2}\sum_i \sigma_x^{(i)} + \frac{B(t)}{2}\left(\sum_i h_i \sigma_z^{(i)} + \sum_{i>j} J_{ij} \sigma_z^{(i)} \sigma_z^{(j)}\right), \qquad (1)$$

where $\sigma_x^{(i)}$ and $\sigma_z^{(i)}$ are $x$ and $z$ Pauli matrices acting on qubit $i$, respectively [10,13,14]. The parameter $h_i$ is self-bias on qubit $i$ and $J_{ij}$ is a coupling constant between qubits $i$ and $j$. In general, $h_i$ and $J_{ij}$ are dimensionless programmable input parameters that define the problem to be solved. The tunable coupler is utilized to tune the intensity of $J_{ij}$. Time-dependent coefficients $A(t)$ and $B(t)$ are the energies that determine the annealing schedule. The second term with $B(t)$, the problem Hamiltonian, is designed so that its ground state(s) correspond to the problem's solution(s). With an appropriate annealing schedule based on the adiabatic theorem, QA provides one of the solutions with high probability.

Here, however, we need a trick to physically realize the general idea as a physical Ising system with a fixed interaction topology like the chimera graph architecture. In cases where the problem Hamiltonian does not fit the fixed topology, we need to construct a new problem Hamiltonian satisfying two conditions: 1) interactions in the new problem Hamiltonian are consistent with the fixed architecture, and 2) (the set of) ground state(s) of the new problem Hamiltonian are "equivalent" to the original one, where we restricted our view to the Hilbert space where the original Hamiltonian lives. (In general, extra qubits are required for such constructions, so that the dimension of the Hilbert space where the new Hamiltonian lives becomes larger than the original one.) Fortunately, such construction of the new Hamiltonian is known to always be possible (theoretically, at least), and automatic software tools for that purpose are available.

Note, however, that the above approach retains some subtle problems, from both computational and practical points of view. The first problem is that performing quantum annealing on the original and the new problem Hamiltonian can have drastically different efficiencies. The above "equivalency" of (a set of) ground state(s) does not imply equivalence of the computational efficiencies. Specifically, since the spectrum structure including excited states is modified, the annealing dynamics can be changed, resulting in drastic differences in the computational efficiencies [16]. To realize the full ability of QA, direct implementation of the original Hamiltonian can be essentially significant. The second problem is scalability. As mentioned above, extra qubits must generally be introduced when constructing the new Hamiltonian. When applying QA to real-world problems in particular, the required number of extra qubits is expected to be significantly increased, placing practical restrictions on the size of problems to which we can apply QA.

With these problems as a motivation, in this article we would like to investigate a different approach from the fixed topology architecture, namely, one by physical implementation of the original problem Hamiltonian itself natively. The idea of "native" implementation of the original problem Hamiltonian has been proposed as an application-specific circuit approach [17]. Following that general idea, in this paper we report experimental demonstrations of native implementations of Boolean logic functions as QA circuits. To enable higher-order operations, a multiplier was also implemented in a specific QA circuit.

## III. Hamiltonian description for application-specific circuits in a quantum annealer

In our approach, $h_i$ is a variable parameter modulated by external flux, while $J_{ij}$ is a parameter fixed by adjusting the overlapping area between qubits $i$ and $j$. Accurate implementation of $h_i$ and $J_{ij}$ components is achieved by precise design of an inductance $L$ of the qubit and a mutual inductance $M$. Values for $L$ and $M$ are extracted from the QA circuit layout using InductEX [18]. The qubit state,

"0" or "1," is defined by the direction of the persistent current. In this study, state 1 indicates clockwise current flowing in the qubit. The readout circuit was composed of a quantum flux parametron (QFP) and a dc superconducting quantum interference device (SQUID). The QFP detects flux due to the persistent current in the qubit and transfers it to the dc-SQUID with an amplification [11]. The dc-SQUID was tuned by applying an external flux to respond to clockwise current in the qubit. We fabricated a QA circuit using superconducting integrated circuit technologies for superconducting flux qubits, providing Nb 4-layers and a Josephson junction (JJ) with a critical current density $J_c$ of 1 µA/µm². The critical current $I_c$ in the qubit was designed as 6.25 µA.

*A. Hamiltonian in the logic gates and their implementation in the QA circuits*

The inset in Fig. 1(a) shows relations between $h_i$ (red circles) and $J_{ij}$ (blue lines) terms in the logic gates. Functionally complete sets of NOR and NAND are expressed as Hamiltonians by alternating the sign of $h_i$ with the same absolute value of coefficients. In AND and OR gates, both signs of the $h_i$ and $J_{ij}$ terms are modified from the NOR and NAND gates. The states of qubit 1 ($Q_1$) and qubit 2 ($Q_2$) correspond to inputs A and B in each logic gate, respectively. The state of qubit 3 ($Q_3$) represents output R. In these configurations, the energy level in each logic gate is minimal at each logic element, as shown in Fig. 1(a). These minimal energies correspond to a ground state in the QA circuit if the Hamiltonian is directly implemented without modification, such as applying an extra penalty.

NOR and NAND gates are implemented in the same QA circuit with 3-qubits. Direct implementations of the logical Hamiltonians are possible in the QA circuit, as shown in Fig. 2(a). In this circuit, $h_i$ terms are modulated by applying flux through self-bias currents (red dashed rectangle). Mutual inductance between the self-bias current and the qubit was designed as 31 pH. $J_{12}$, $J_{13}$, and $J_{23}$ were implemented as M of 10, 22, and 22 pH, respectively. The L of each qubit was designed as 110 pH. Accuracy of values in L and M were experimentally confirmed using a circuit manufactured separately from the QA circuit.

The other QA circuit was prepared for AND and OR gates, where current direction at the overlapping area between qubits 1 (2) and 3 (3) differed from the circuit for NOR and NAND gates (not shown, but with structures nearly the same as those in Fig. 2(a)).

*B. Hamiltonian in the multiplier and its QA circuit*

Figure 1(b) shows the Hamiltonian configuration of the multiplier and its energy level. The multiplier consisted of six qubits, corresponding to X and Y for inputs, Z and D for carry-in, C for carry-out, and S for summation. In this configuration, sixteen combinations of (X, Y, Z, D, C, S) take minimal energy.

Figure 2(b) shows the QA circuit of the multiplier where the logical Hamiltonian is directly implemented. The mutual inductances between self-bias current and the qubits were designed as 31 pH. From analysis of L based on the layout of the multiplier using InductEX, L values for $Q_1$–$Q_6$ were 278, 278, 284, 287, 277, and 300 pH, respectively. Relations between $J_{ij}$ terms illustrated in Fig. 1(b) were implemented by tuning the overlapping area between qubits and analytically estimated as shown in Table 1. Accuracy of estimated values in L and M were experimentally confirmed using a circuit manufactured separately from the QA circuit.

### III. Verification of the functionally complete set

Since ground states of the Hamiltonian appear after QA processing, combinations of qubit states are considered to be consistent with each logic function. Each qubit state is evaluated by applying a time-dependent transverse field to the QA. The annealing effect is controlled by the sweeping time ($T_a$) of the field. Figure 3(a) shows the experimental setup in each qubit consisting of the QA circuit. Arbitrary wave generators are utilized to apply flux with an accurate time schedule. $I_{trans}$, $I_{QFP}$, $I_{bias\_sq}$, $I_{flux\_sq}$, and $I_h$ are current for the transverse field, flux-injection current for the QFP, drive current for the dc-SQUID, modulation current for flux detection in the dc-SQUID, and self-bias current in the qubit, respectively. An annealing schedule is controlled by the $I_{trans}$ rise time. Maximum amplitude of the $I_{trans}$ corresponds to injection of the flux quantum $\Phi_0$ (2.07×10$^{-15}$ Wb) to the rf-SQUID in the qubit. After the $\Phi_0$ injection, the QFP and the dc-SQUID are activated as shown in Fig. 3(b). State detection of the qubit is carried out during 1.0-1.1 ms. The state is typically evaluated over 10,000 iterations.

## A. Verification of functionally complete set

Figure 4(a) shows a histogram for NOR at currents ($I_{h1}$, $I_{h2}$, $I_{h3}$) = (-1.99, -1.42, -4.50) [μA] over 10,000 iterations in experiments conducted at 4.2 K and 10 mK. In both cases, $T_a$ was 100 μs. NOR logic elements "001", "010", "100", and "110" appeared. Undesirable elements were observed in experiments at 4.2 K, but suppressed at 10 mK. The thermal energies at 4.2 K and 10 mK are estimated to be $5.8 \times 10^{-23}$ and $1.38 \times 10^{-25}$ J, respectively. The order of energy in the potential of the rf-SQUID in the qubit around its bottom is approximately $4.9 \times 10^{-22}$ J. Thermal fluctuation might thus be largely affected in experiments at 4.2 K.

Figure 5(a) shows a typical demonstration of random NOR operations at currents ($I_{h1}$, $I_{h2}$, $I_{h3}$) = (-1.99, -1.42, -4.50) [μA]. Combinations of signals detected by the dc-SQUID ($V_A$, $V_B$, $V_C$) show one of the NOR elements. Figure 5(b) shows random operation in NAND when the self-bias sign is inverted. Figures 5(c) and 5(d) show OR and AND operations. The logic function is intentionally producible by applying an appropriate offset current α. If we demonstrate the logic "001" in NOR, self-bias currents of ($I_{h1}'$, $I_{h2}'$, $I_{h3}'$) = ($I_{h1}$ - α, $I_{h2}$ - α, $I_{h3}$) are applied. Figures 5(e) and 5(f) exhibit intentional generation of logics "001" and "110" in NOR, respectively. These results indicate the possibility of a versatile computing system using QA with combinations of this functionally complete set.

## B. Demonstration of multiplication

Figure 4(b) shows a histogram of the multiplier at current ($I_{h1}$, $I_{h2}$, $I_{h3}$, $I_{h4}$, $I_{h5}$, $I_{h6}$) = (-0.28, -0.23, -0.45, -0.6, 0.37, 0.42) [μA] over 10,000 iterations. The histogram was investigated by modulating $T_a$ between 1 and 1000 μs. In both $T_a$, all sixteen candidate logic elements, which are at the minimum energy level in Fig. 1(b), are observed. In Fig. 1(b), "other" means incorrect combinations of ($X$, $Y$, $Z$, $D$, $C$, $S$) in multiplier responses. The ratio of "other" components in the histogram was rather large. The main reason is the existence of a local minimum in an energy potential of each qubit. To avoid generation of local minima in energy potentials, the value of a dimensionless factor $\beta_L = 2\pi L I_c / \Phi_0$, where $L$ is the inductance, should be less than 8 [19]. Since the $\beta_L$ of the multiplier was 10.8, an unintentional trap to the local minimum occurred during the annealing process, resulting in incorrect multiplier responses. In future work, we will reduce both the $L$ and $I_c$ values in each qubit to control the $\beta_L$ value to less than 8. Fortunately, all sixteen candidate logic elements were obtained, indicating the possibility of intentional multiplication by applying offset current a to this self-bias condition.

A superconductor simulation program with an integrated circuit emphasis (SPICE) model was established in the multiplier by considering behavior of the qubit including influences of thermal and quantum noise in the experiment at 10 mK. $L$ and $M$ parameters extracted from the QA circuit layout were utilized in the circuit model. Owing to a time constraint, $T_a$ settled in 1 μs. After annealing, the QFP is activated. Later, dc-SQUID is validated from 1.04-1.09 ms for a read operation. The model was analyzed using a Josephson integrated circuit simulator (JSIM) [20]. Based on the current ($I_{h1}$, $I_{h2}$, $I_{h3}$, $I_{h4}$, $I_{h5}$, $I_{h6}$) = (-0.22, -0.36, -1.00, -1.00, 0.80, 0.60) [μA], by which all sixteen candidate logic elements were obtained by JSIM analysis, intentional logic elements were calculated with α = 2.0 μA. Here, if we calculate input combinations of ($X$, $Y$, $Z$, $D$) = (0, 0, 1, 0), self-bias currents of ($I_{h1}'$, $I_{h2}'$, $I_{h3}'$, $I_{h4}'$, $I_{h5}'$, $I_{h6}'$) = ($I_{h1}$- α, $I_{h2}$- α, $I_{h3}$+ α, $I_{h4}$- α, $I_{h5}$, $I_{h6}$) are applied. Figure 6 shows typical responses of the multiplier. Correct responses were obtained in each calculation. Figure 7 shows the same combinations of logic elements obtained by an experiment at 10 mK with $T_a$ = 100 μs. The results exhibit the same response as the JSIM analysis.

## C. Confirmation of functioning in a scalable factorization unit

Note that backward computation (finding input configurations based on a given output with respect to the logic function) is possible in QA systems [17, 21]. This basic idea provides the possibility of factorization using a multiplier. A scalable factorization circuit is possible by assembling the multiplier. In the backward calculation, the initial states in qubits $Q_5$(C) and $Q_6$(S) are tuned by considering α when the QA is performed. Figure 8 shows JSIM analysis of factorization in the multiplier. Although the wrong response was generated, due to the short $T_a$ (1 μs) caused by limited simulation resources, all possible elements appeared in each factorization. These features were confirmed by an experiment at 10 mK (Fig. 9). Figure 10 shows offset current a dependencies of success probability in the factorization. Increasing the offset current improved accuracies. Though the multiplier in this work had local minima in its energy landscape owing to a large $β_L$, demonstrations of factorization were successfully achieved. This indicates the application-specific QA circuit has a certain level of tunability by applying offset current.

## IV. Domain-specific quantum computing

Domain-specific computing, where a specialized processing unit is selected based on a task feature during the entire execution process, has been a major trend in classical computing using von Neumann architectures [22-24]. Recently, system architectures with heterogeneous accelerators composed of classical computers and quantum computers have been proposed [25]. Figure 11(a) shows one candidate hybrid architecture, where each quantum computer is connected on the network. Here, the host computer selects the most appropriate computer to calculate the task. Figure 11(b) shows one early prototyping of this domain-specific quantum computing (DSQC), where processing units have two different QA circuits on the same chip and the host computer is connected via a network. Here, the two types of QA circuits correspond to specialized processing units for different tasks. As Fig. 11(c) shows, two processing units, the NOR (QA circuit 1) and the multiplier (QA circuit 2) were implemented on the same chip. In our experimental setup, the control unit can switch operations between the NOR and the multiplier. An operator can select the processing unit from a host computer connected with the control unit via a network. We consider this as a proof-of-concept demonstration in early prototyping of DSQC and hybrid computing architectures.

## V. CONCLUSION

We investigated a novel approach unlike topology-fixed-architectures such as the D-Wave Systems chimera graph, namely, native physical implementation of the original problem Hamiltonian itself. As a superconducting integrated circuit, a problem Hamiltonian whose set of ground states is consistent with a given truth table is implemented for the QA circuit with no redundant qubits. We implanted a functionally complete set and multiplication as examples of the truth table. We applied a static coupler for implementation of the $J_{ij}$ term. Owing to precise design of $L$ and $M$, the original problem Hamiltonian was accurately expressed as a QA circuit. NOR(OR) and NAND(AND) were identically implemented in the same QA circuit composed of three superconducting flux qubits. The function was selected by switching the sign of the self-bias current. By applying the offset current, each logic function was intentionally reproduced. This demonstration indicated the possibility of a versatile computing system using QA with combinations of this functionally complete set. We also demonstrated a QA circuit for a multiplier comprising six qubits. In the forward process, where the offset currents were adopted in the input qubits, multiplication was successfully confirmed both in the JSIM analysis and in experiments at 10 mK. One advantage to utilizing a QA system as compared with classical computing is the possibility of backward computation, where the offset currents were adopted in the output qubits. We successfully demonstrated the backward process in a multiplier corresponding to factorization, both in JSIM analysis and experiments at 10 mK. Functionality is easily expandable by assembling units, resulting in a scalable system that is not readily achievable by conventional QA systems. The QA circuits demonstrated in this paper were implemented on the same chip and selectively operated from the control unit. Since we can access the control unit from the host computer via a network, our demonstration can be considered as a proof-of-concept demonstration for early prototyping of DSQC and hybrid computing architectures. Combinations of functionally complete sets will realize versatile processing units, while expanding the multipliers would be effective for prime factorization units.

## ACKNOWLEDGMENT

The authors thank K. Inomata for his dedicated support in the experiments, and Y. Araga and K. Kikuchi for their heartfelt cooperation.

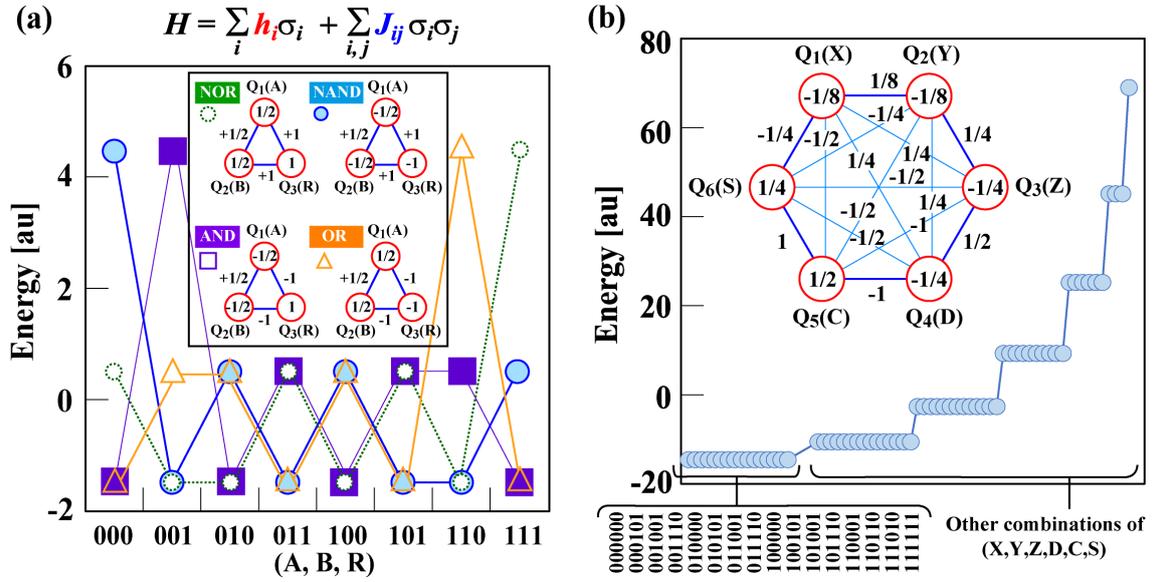

**FIGURE 1.** Energy of each state in (a) logic gates and (b) a multiplier unit. The inset in Fig. 1(a) shows relations between $h_i$ (red circles) and $J_{ij}$ (blue lines) terms of Hamiltonians in each logic gate.

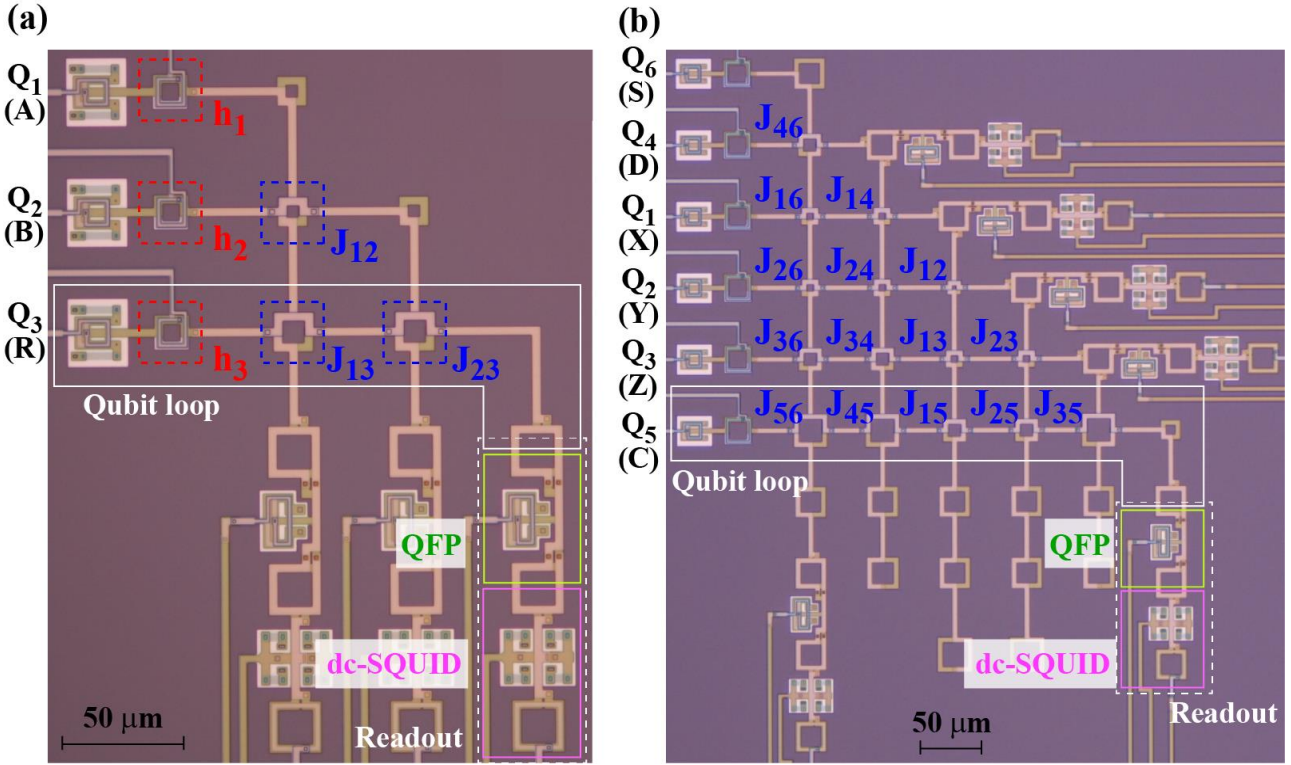

**FIGURE 2.** Photograph of QA circuits for (a) the logic gate and (b) the multiplier unit. $Q_i$ represents the superconducting flux qubits. The logic gate and the multiplier respectively comprise 3 and 6 qubits. $J_{ij}$ terms are implemented by tuning the overlapping area between qubits $i$ and $j$. The qubit state is detected by a readout consisting of the QFP and the dc-SQUID. In Fig. 2(a), $(J_{12}, J_{23}, J_{31})$ correspond to $(0.5, 1.0, 1.0)$ and $(0.5, -1.0, -1.0)$ in the QA circuit for NOR(NAND) and OR(AND), respectively.

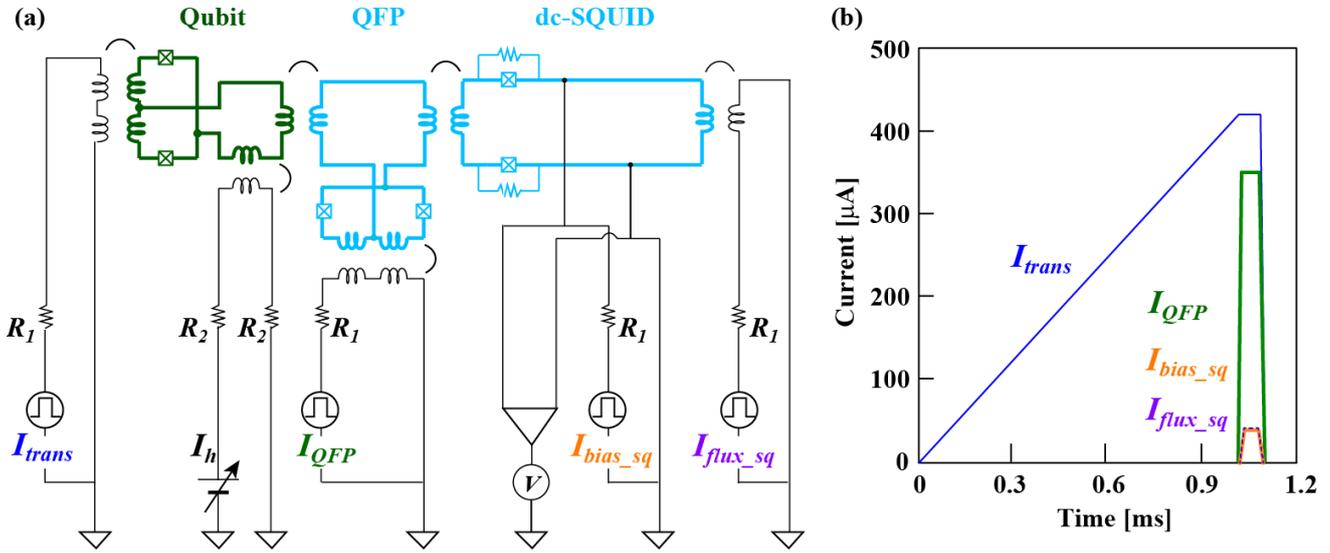

**FIGURE 3.** (a) Experimental setup for an individual qubit of the QA circuit and a schedule of applied currents. $R_1$ and $R_2$ were 10 kΩ and 1 MΩ, respectively.

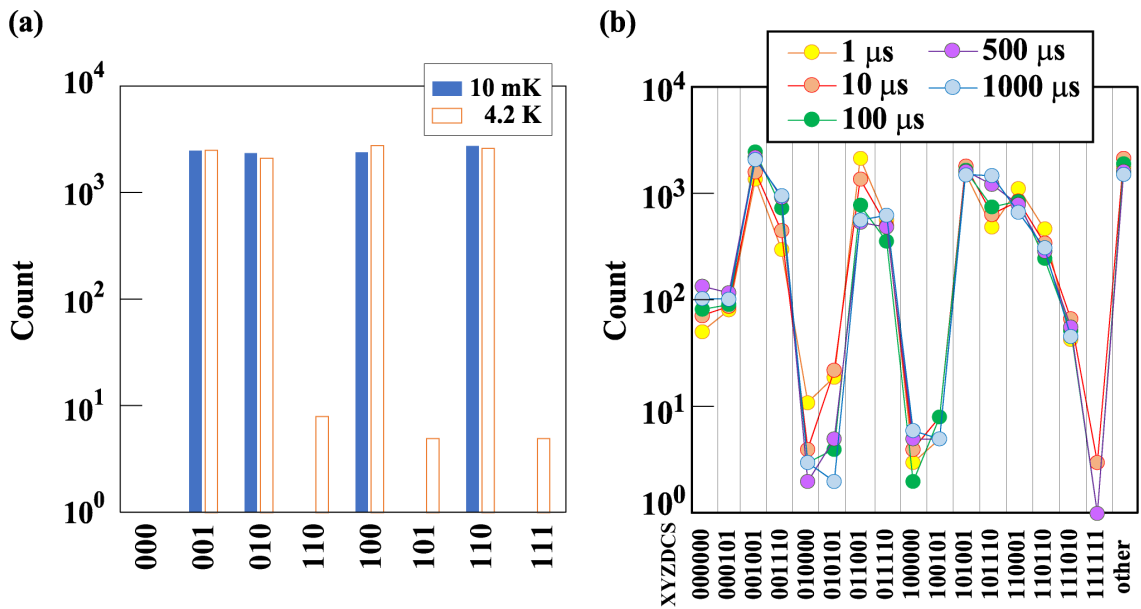

**FIGURE 4.** Histograms of (a) NOR and (b) the multiplier unit at the base condition for self-bias currents.

TABLE I

IMPLEMENTED MUTUAL INDUCTANCES CORRESPONDING TO $J_{ij}$ TERMS

| Mutual inductance [pH] | | | | | |
|---|---|---|---|---|---|
| $J_{12}$ | -5.6 | $J_{23}$ | -9.9 | $J_{35}$ | 44.5 |
| $J_{13}$ | -9.7 | $J_{24}$ | -11.0 | $J_{36}$ | 20.4 |
| $J_{14}$ | 10.0 | $J_{25}$ | 20.7 | $J_{45}$ | 44.0 |
| $J_{15}$ | 20.7 | $J_{26}$ | 11.5 | $J_{46}$ | 23.0 |
| $J_{16}$ | 10.1 | $J_{34}$ | -20.3 | $J_{56}$ | -43.4 |

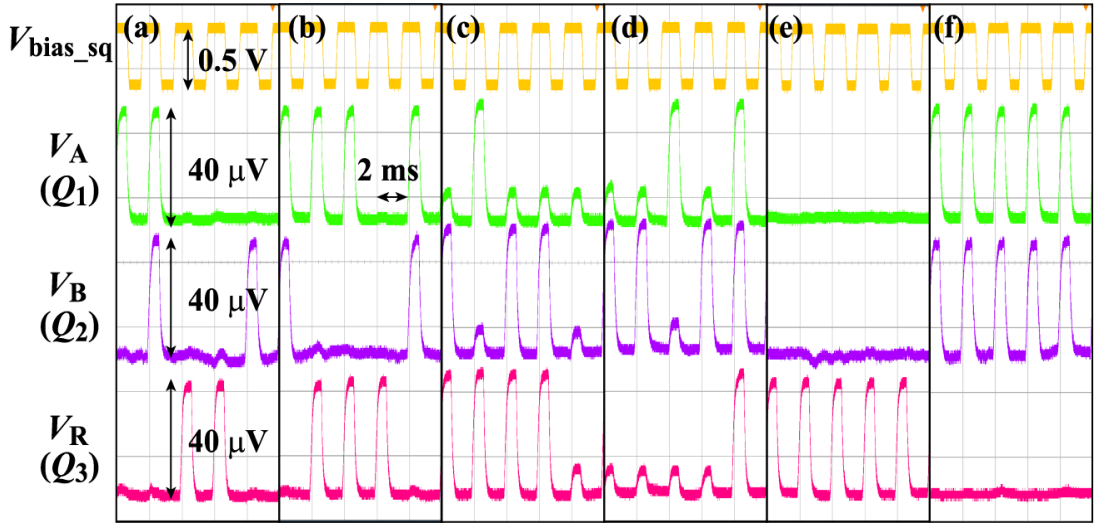

**FIGURE 5.** Responses of the dc-SQUID in (a) NOR, (b) NAND, (c) OR, and (d) AND at 10 mK. Experiment was carried out with an annealing time ($T_a$) of 100 µs. An arbitrary element is obtained by applying offset current $\alpha$. (e) 001 and (f) 110 in NOR operation with $\alpha = 0.5$ µA.

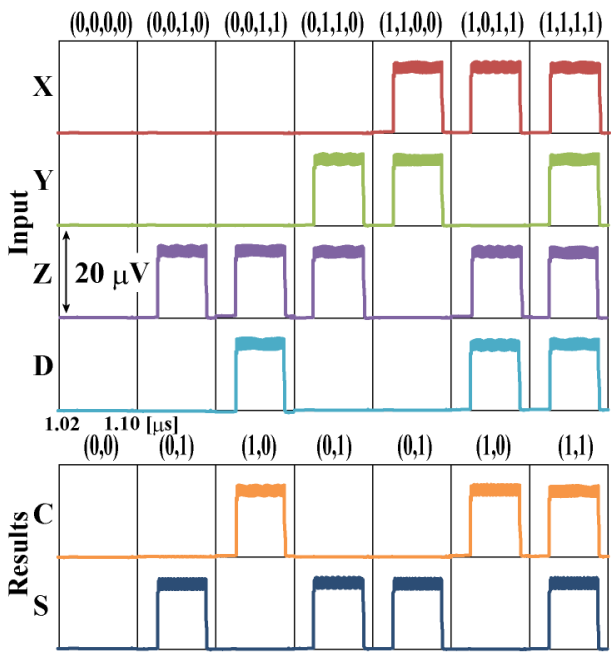

**FIGURE 6.** Typical responses of a multiplication during read in the multiplier unit calculated by JSIM with $T_a = 1$ µs. In these calculations, self-bias currents in $Q_1$, $Q_2$, $Q_3$, and $Q_4$ are tuned by considering $\alpha$ during QA. DC-SQUID is validated from 1.04 µs to 1.09 µs. Here, state 1 in the qubit is detected.

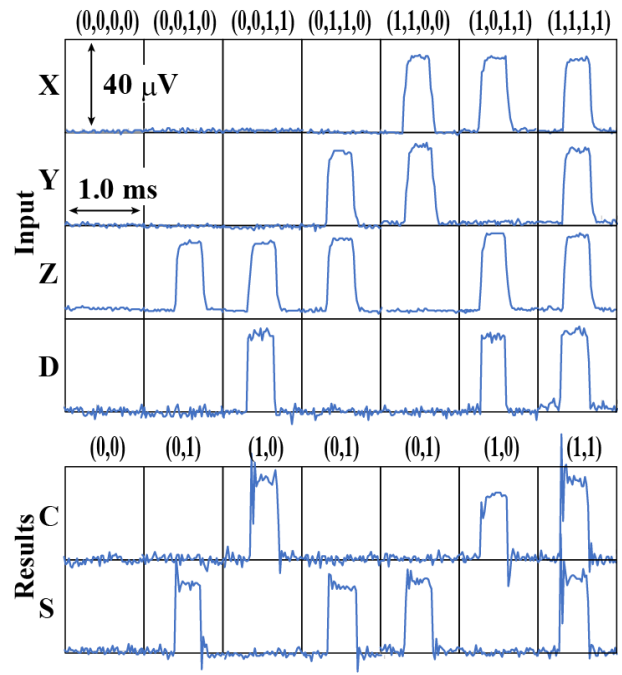

**FIGURE 7.** Typical responses in the multiplier at 10 mK (forward process). Experiments were carried out with $T_a = 100$ µs. In these experiments, self-bias currents in $Q_1$, $Q_2$, $Q_3$, and $Q_4$ are tuned by considering $\alpha$ during QA.

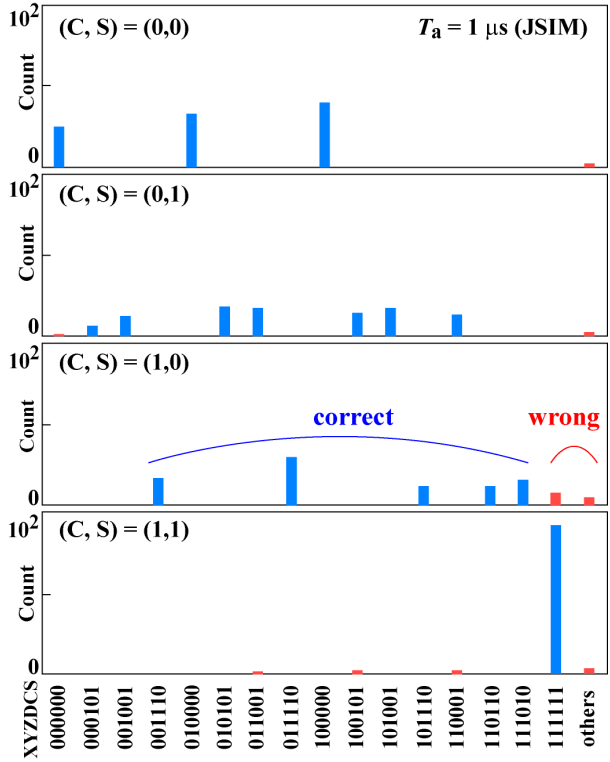

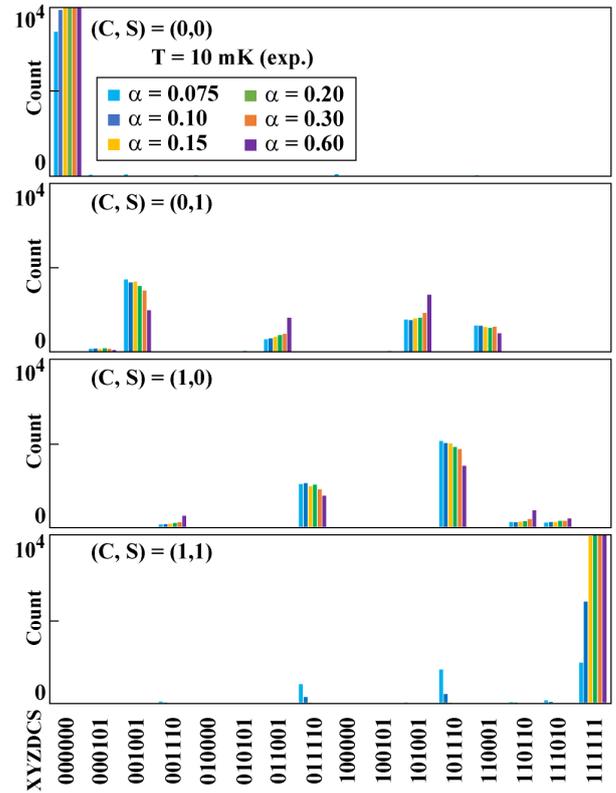

FIGURE 8. SPICE analysis of factorization using the multiplier unit with $T_a = 1$ μs. Here, self-bias currents of $Q_5$ and $Q_6$ are tuned by considering α.

FIGURE 9. Demonstration of factorization using the multiplier at $T = 10$ mK (backward process). Experiment was carried out with the $T_a = 100$ μs. In these experiments, self-bias currents of $Q_5$ and $Q_6$ are tuned by considering α.

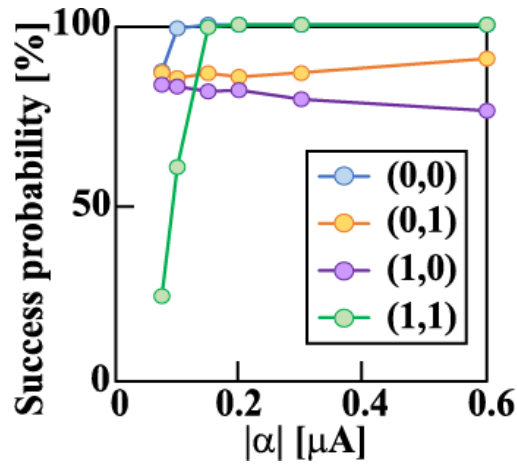

FIGURE 10. Offset current α dependence of the success probability in the factorization using the multiplier unit.

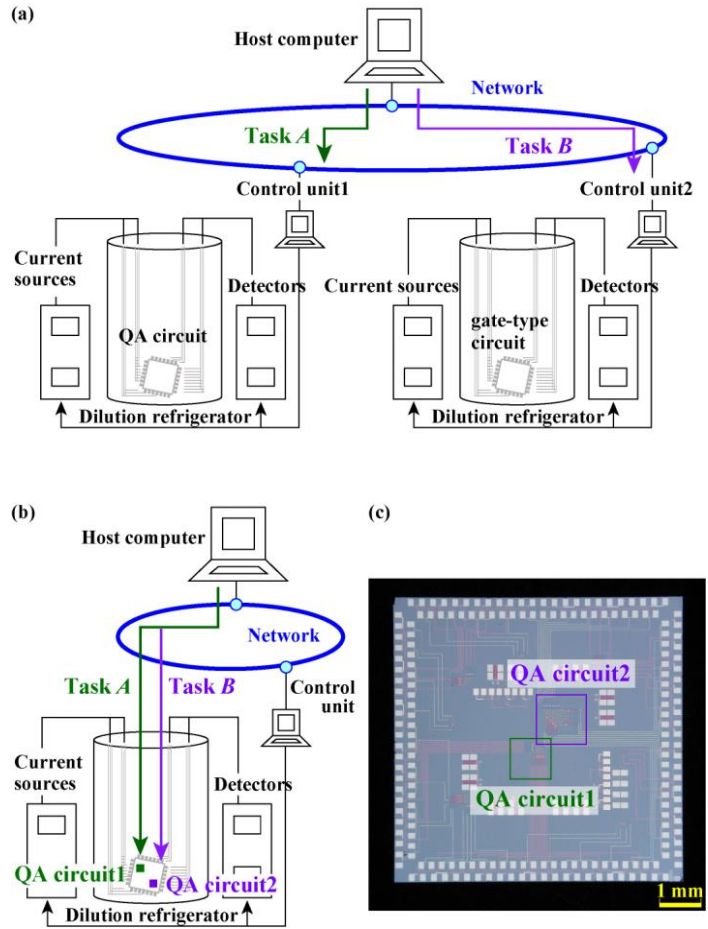

**FIGURE 11.** (a) Schematic for one candidate hybrid architecture comprising a QA circuit and a gate-type quantum circuit. (b) Schematic for one early prototyping of the hybrid architecture comprising domain-specific quantum computing. (c) Photograph of a QA circuit with the NOR (QA circuit 1) and the multiplier (QA circuit 2) on the same chip. In our experiment, the control unit could switch operations between NOR and the multiplier.